\documentclass[twocolumn,showpacs,aps,floatfix,longtable,pre]{revtex4}

\usepackage{graphicx}
\usepackage{bm}

\begin{document}

\title{Ising antiferromagnet on the Archimedean lattices}
\author{Unjong \surname{Yu}}
\affiliation{Division of Liberal Arts and Sciences \& Department of Physics and Photon Science, Gwangju Institute of Science and Technology, Gwangju 500-712, South Korea}
\email[E-mail me at: ]{uyu@gist.ac.kr}

\begin{abstract}
Geometric frustration effects were studied systematically with the Ising antiferromagnet
on the 11 Archimedean lattices using the Monte-Carlo methods.
The Wang-Landau algorithm for static properties (specific heat and residual entropy) and
the Metropolis algorithm for a freezing order parameter were adopted.
The exact residual entropy was also found.
Based on the degree of frustration and dynamic properties, ground states of them were determined.
The Shastry-Sutherland lattice and the trellis lattice are weakly-frustrated and
have two-dimensional and one-dimensional long-range-ordered ground states, respectively.
The bounce, maple-leaf, and star lattices have the spin ice phase. The spin liquid phase
appears in the triangular and kagom\'e lattices.
\end{abstract}



\maketitle

\section{Introduction}

The Ising model \cite{Ising25} has played a crucial role to understand the phase transition and magnetic ordering.
There exists a phase transition at finite temperature from a high-temperature disordered phase
into a low-temperature ordered phase in a lattice of two and three dimension,
if it is not frustrated \cite{Peierls36,Onsager44,Wannier50,Kano53}.
With frustration, the Ising model has various ground states such as long-range-ordered
phase, spin glass, spin ice, and spin liquid phase \cite{Balents10,Diep}.
Frustration can be induced in two ways: by disorder (spin glass)
or by geometry (geometrically frustrated systems) \cite{Ramirez94}.
In this paper, effects of the geometric frustration are investigated systematically
for the antiferromagnetic Ising model on the two-dimensional Archimedean lattices.

\begin{figure}[b]
\includegraphics[width=8.4cm]{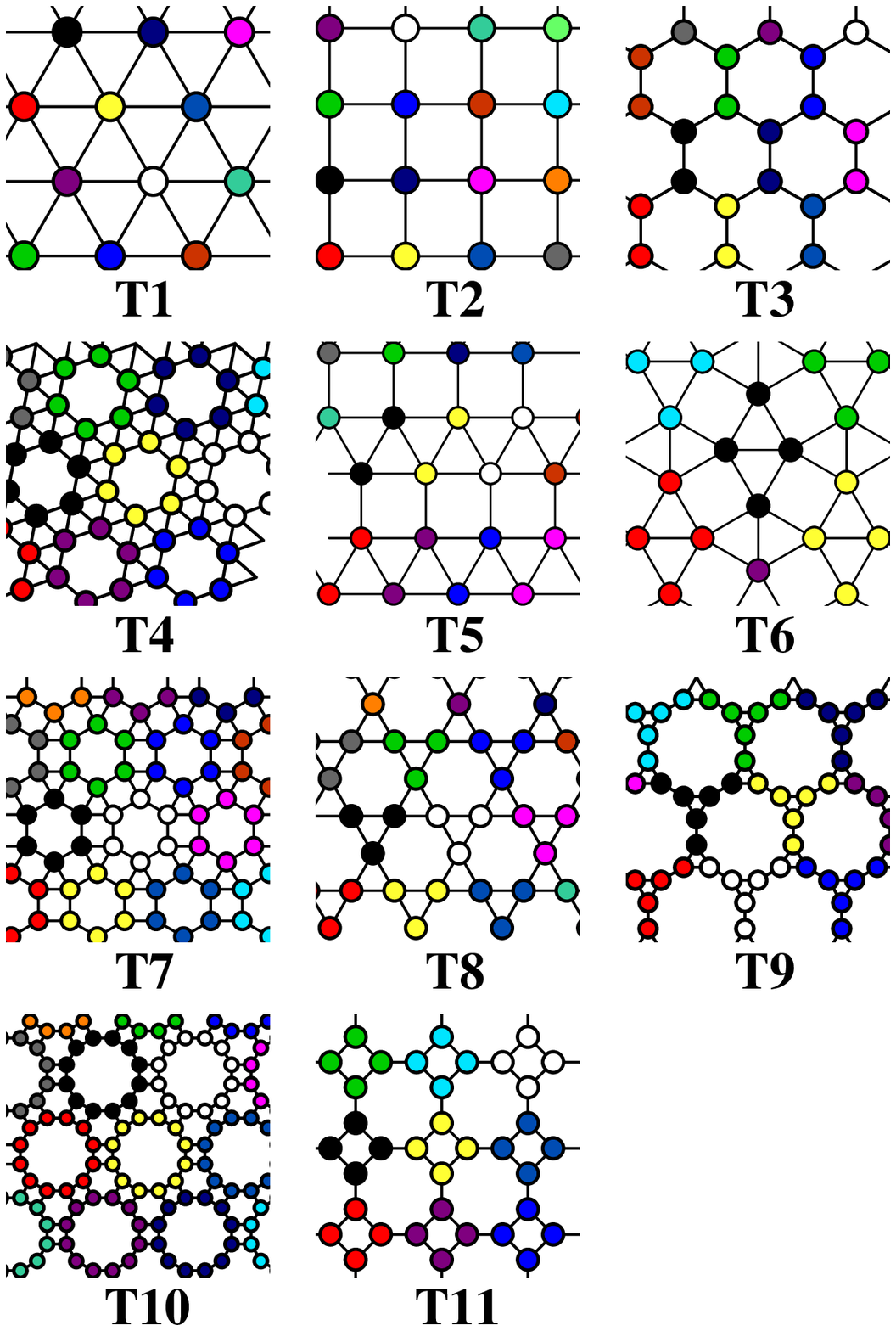}
\caption{(Color online) The 11 Archimedean lattices. Lattice points in the same unit cell
are represented by circles of the same color (gray scale).}
\label{Arch_fig}
\end{figure}

An Archimedean lattice, also called as a uniform tiling, is a two-dimensional lattice of regular polygons
in which all vertices are topologically equivalent.
It is known that Kepler proved that there exist only 11 Archimedean lattices \cite{Grunbaum87},
which are listed in Fig.~\ref{Arch_fig} and Table~\ref{Arch_table}.
They are important not only in mathematics but also in materials science,
because most of the lattices have corresponding natural material systems \cite{Richter04,Zheng14}.
Since all vertices are in the same topological environment,
they can be labeled by the sequence of the polygons surrounding a vertex \cite{Grunbaum87}.
For example, a square and a triangular lattices are represented by $(4^4)$ and $(3^6)$, respectively.
We follow the naming convention of Ref.~\onlinecite{Richter04}, where 
homopolygonal lattices (T1, T2, T3) are followed by heteropolygonal lattices (T4, $\cdots$, T11).
Due to the simplicity, the Archimedean lattice is a good starting point to study geometric frustration.
With the antiferromagnetic nearest-neighbor interaction,
4 lattices are bipartite and unfrustrated, and the other 7 lattices are frustrated. 

There have been systematic studies about percolation on the Archimedean lattices \cite{Suding99,Neher08}.
As for the magnetic models, the ferromagnetic Ising model \cite{Codello10}
and antiferromagnetic quantum Heisenberg model \cite{Richter04,Farnell14} were studied.
In this paper, we report detailed study of the antiferromagnetic Ising model on the Archimedean lattices.
Specific heat, exact residual entropy, and freezing order-parameter are obtained to
identify the ground state. Finite temperature phase transitions
in weakly frustrated lattices are also discussed.

\begin{table*}[b]
\caption{\label{Arch_table}Name, number of lattice points per basis ($B$), bipartiteness (Bip.), coordination number ($z$),
antiferromagnetic ground state energy per bond ($E_g^{\rm AF}$), 
ferromagnetic and antiferromagnetic transition temperatures ($T_{c}$),
exact residual entropy ($S_0$), and ground state for each Archimedean lattice.
Ferromagnetic transition temperatures are from Ref.~\protect\onlinecite{Codello10}, where exact values are given.
Results of this work is in bold.
}
\begin{ruledtabular}
\begin{tabular}{ccccccccccc}
 & Name & & $B$ & Bip. & ~$z$~ & $E_g^{\rm AF}$ & $T_{c}^{\rm F}$ & $T_{c}^{\rm AF}$ & $S_0$ & Ground state\\
\hline
T2 & Square      & ($4^4$)       & 1 & Y & 4 & $-1$      & 2.2692 & $=T_{c}^{\rm F}$ & $\log(2)$  & Long-range-order      \\
T3 & Honeycomb   & ($6^3$)       & 2 & Y & 3 & $-1$      & 1.5187 & $=T_{c}^{\rm F}$ & $\log(2)$  & Long-range-order      \\
T11& CaVO (CaV$_4$O$_9$)\footnotemark[1] & ($4,8^2$) & 4 & Y & 3 & $-1$ & 1.4387 & $=T_{c}^{\rm F}$ & $\log(2)$  & Long-range-order      \\
T10& SHD\footnotemark[2]  & ($4,6,12$)    & 12 & Y & 3 & $-1$      & 1.3898 & $=T_{c}^{\rm F}$ & $\log(2)$  & Long-range-order      \\
\hline
T6 & SSL\footnotemark[2]  & ($3^2,4,3,4$) & 4 & N & 5 & $-0.6$ & 2.9263 & {\bf 1.261(1)} & $\log(2)$  & Long-range-order      \\
T5 & Trellis     & ($3^3,4^2$)   & 2 & N & 5 & ${\bf -0.6}$    & 2.8854 & {\bf 0.19(2)}  & ${\bf log(2)} \bm{L}$& {\bf Long-range-order}\footnotemark[3]\\
\hline
T7 & Bounce      & ($3,4,6,4$)   & 6 & N & 4 & ${\bf -0.6667}$ & 2.1433 & -                & ${\bf 0.0538} \bm{N}$ & {\bf Spin ice}              \\
T4 & Maple leaf  & ($3^4,6$)     & 6 & N & 5 & ${\bf -0.4667}$ & 2.7858 & -                & ${\bf 0.0538} \bm{N}$ & {\bf Spin ice}              \\
T9 & Star\footnotemark[1]        & ($3,12^2$)& 6 & N & 3 & ${\bf -0.5556}$ & 1.2315 & -    & ${\bf 0.2509} \bm{N}$ & {\bf Spin ice}              \\
\hline
T1 & Triangular  & ($3^6$)       & 1 & N & 6 & ${\bf -0.3333}$ & 3.6410 & -                & ${\bf 0.3231} \bm{N}$ & {\bf Spin liquid}           \\
T8 & Kagom\'e    & ($3,6,3,6$)   & 3 & N & 4 & ${\bf -0.3333}$ & 2.1433 & -                & ${\bf 0.5018} \bm{N}$ & {\bf Spin liquid}           \\
\end{tabular}
\footnotetext[1]{T11 and T9 are also called bathroom tile and expanded kagom\'e, respectively.}
\footnotetext[2]{SSL and SHD are abbreviations for Shastry-Sutherland lattice \protect\cite{SSL} and square-hexagonal-dodecagonal, 
                 respectively.}
\footnotetext[3]{Long-range-order only in one direction.}
\end{ruledtabular}
\end{table*}

\section{Model and method}
The Ising model studied in this work is represented by the following Hamiltonian.
\begin{eqnarray}
H = - J \sum_{\langle i,j \rangle} S_{i} S_{j}
\label{Eq:Ising}
\end{eqnarray}
The spin at the $i$-th site $S_{i}$ may take the values of $+1$ or $-1$, only.
The summation runs for all the nearest neighbor pairs, excluding double counting.
The coupling constant $J$ is set to $-1$ and $|J|/k_B$ is used as an energy unit,
where $k_B$ is the Boltzmann constant.
Negative $J$ means an antiferromagnetic interaction.

The calculation was done for parallelepiped lattices with number of sites $N=B\times L\times L$
with the periodic boundary condition. $B$ is the number of sites per unit cell and
$L$ is the linear size. Without frustration, the most efficient algorithm
for the Ising model is the combination of the Wolff cluster-update \cite{Wolff89}
and the histogram reweighting \cite{Ferrenberg88,Ferrenberg89}.
In the ferromagnetic (antiferromagnetic) case, a cluster is made by adding
nearest neighbors of the same (opposite) spin with a probability $P=1-\exp(-2/T)$.
When a cluster is completed, all spins of the cluster are flipped.
The cluster-update is not efficient away from the critical temperature,
where the cluster size is too small (at high temperature) or too large
(at low temperature). Close to the critical temperature,
the cluster size is moderate and the critical slowing down can be eliminated.
The information of energy and magnetization distribution at a fixed temperature
is used by the histogram method to calculate thermodynamic quantities
at any temperature near the simulation temperature \cite{Yu15}.
With frustration, however, cluster-update methods do not work,
because the cluster may include the whole lattice and
cluster size becomes too large even without a phase transition.
Therefore, we used the Wang-Landau algorithm \cite{Wang01} in this study.
While conventional Monte-Carlo methods such as the Metropolis \cite{Metropolis53}
and the Wolff algorithm \cite{Wolff89}
perform a simulation within the canonical ensemble at a fixed temperature,
the Wang-Landau algorithm calculates the density of states (DOS)
as a function of energy $\rho(E)$ directly by the random walk
with the transition probability $P(i\rightarrow j) = \min \{1 , \rho(E_i) / \rho(E_j)\} $ in the whole energy space.
The DOS is adjusted by $\rho(E)\rightarrow f_n \, \rho(E)$ at each step resulting in a flat histogram in the energy space.
The simulation continues until the histogram becomes flat, when its standard deviation is less than 4\% of its average.
Then a new iteration is performed with a smaller value of $f_{n+1} = \sqrt{f_n}$.
The whole simulation begins with an initial multiplier $f_0 = e$ and stops when $f_n$ becomes close enough to 1: $f_n < \exp(10^{-10})$.
The error in DOS depends on many factors, but can be assumed to be the same order as $\log(f_n)$ if the calculation is proper.
Since the random walk is in the whole energy space, it is not stuck to a metastable state and can be applied to any system even with
frustration or first-order transition. From the obtained $\rho(E)$, the average energy $\langle E \rangle_T$
and the specific heat $c(T)$ can be calculated easily as a function of temperature $T$:
\begin{eqnarray}
\langle E \rangle _T = \frac{\sum_{i} E_i \, \rho(E_i) \, e^{-E_i/T}}{\sum_{i} \rho(E_i) \, e^{-E_i/T}} \\
\langle E^2 \rangle _T = \frac{\sum_{i} E_i^2 \, \rho(E_i) \, e^{-E_i/T}}{\sum_{i} \rho(E_i) \, e^{-E_i/T}} \\
c(T) = \frac{ \langle E^2 \rangle_T - {\langle E \rangle}_T ^2}{T^2}
\label{Wang_Landau}
\end{eqnarray}
Since the Wang-Landau algorithm can give only static information,
we also used the Metropolis algorithm to study fluctuation around
     an equilibrium state or among equilibrium states.

\section{Results and discussion}


\begin{figure}[b]
\includegraphics[width=8.4cm]{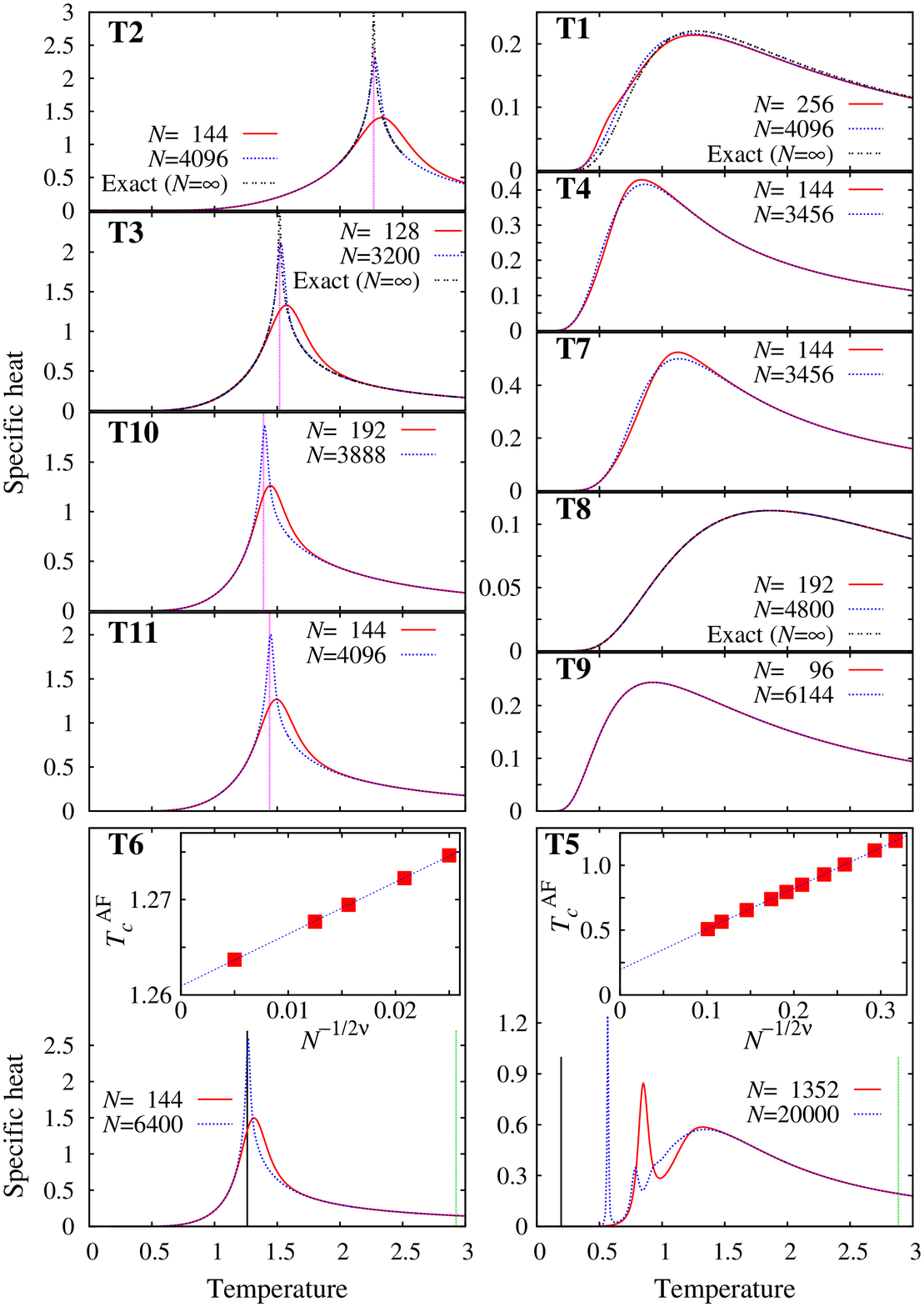}
\caption{(Color online) Specific heat as a function of temperature in the Archimedean lattices.
Upper left panel is for bipartite lattices (T2, T3, T10, and T11)
with long-range-ordered ground states.
Vertical dashed lines represent exact $T_c^{\mathrm{F}} = T_c^{\mathrm{AF}}$.
Upper right panel is for strongly frustrated lattices (T1, T4, T7, T8, and T9).
In the case of weakly frustrated lattices (T6 and T5),
the antiferromagnetic critical temperatures ($T_c^{\mathrm{AF}}$) were 
calculated by the finite-size-scaling shown in the inset with the critical exponent
$\nu=1$ and $\nu=2.3(1)$ for T6 and T5, respectively.
They are $T_c^{\mathrm{AF}} = 1.261(1)$ for T6 and $T_c^{\mathrm{AF}} = 0.19(2)$ for T5.
The two vertical lines represent $T_c^{\mathrm{F}}$ (green, right) and $T_c^{\mathrm{AF}}$ (black, left).
Exact results of the infinite size lattice are from Ref.~\protect\onlinecite{Onsager44}, 
Ref.~\protect\onlinecite{Wannier50}, and Ref.~\protect\onlinecite{Kano53}
for the square, triangular, and kagom\'{e} lattices, respectively.
}
\label{Cv}
\end{figure}

We calculated the specific heat as a function of temperature using the Wang-Landau algorithm,
as shown in Fig.~\ref{Cv}.
The bipartite lattices (T2, T3, T10, and T11) show a phase transition to the antiferromagnetically
ordered ground state at the same critical temperatures as their ferromagnetic cases,
which are represented by the vertical lines.
For strongly frustrated lattices (T1, T4, T7, T8, and T9),
the specific heat shows only a broad maximum without a critical feature, implying a disordered
ground state. Another important feature is very weak size-dependence of the specific heat,
which can be understood by the absence of the long-range-correlation.

In the case of T6 (SSL), the frustration is weak and an ordered ground state is realized.
However, the transition temperature $T_c^{\mathrm{AF}} = 1.261(1)$ is much lower than
its ferromagnetic one ($T_c^{\mathrm{F}} \approx 2.9263$). The transition temperature
was obtained by the finite-size-scaling of $T_c^{\mathrm{AF}}(N) = T_c^{\mathrm{AF}} + A N^{-1/2\nu}$
with the critical exponent $\nu = 1$ of the two-dimensional Ising universality class
and a lattice-dependent constant $A$ \cite{FSS}.
The trellis lattice (T5) shows
both of size-independent broad peak and size-dependent diverging peak.
The latter represents a phase transition at $T_c^{\mathrm{AF}} = 0.19(2)$
with $\nu = 2.3(1)$. Large value of $\nu$ implies the one-dimensional character of the ordering.


\begin{figure}[b]
\includegraphics[width=8.2cm]{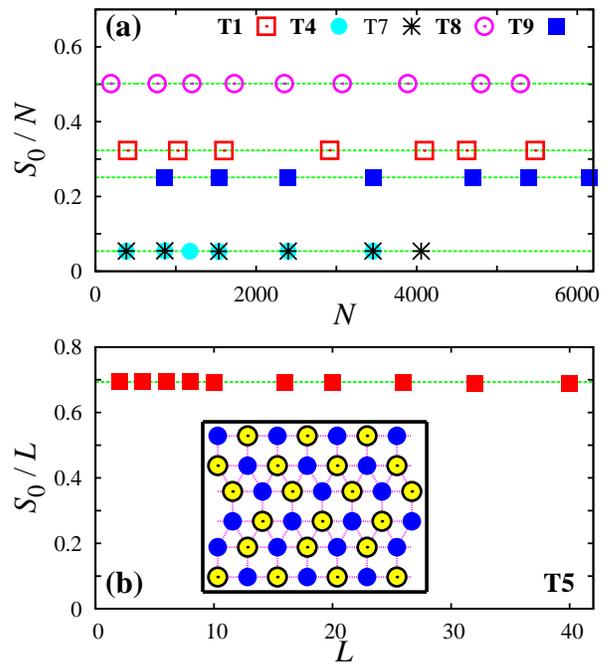}
\caption{(Color online) (a) Residual entropy $S_0$ per lattice point $N$ calculated by the Wang-Landau method
for strongly frustrated lattices without long-range-ordered ground state.
Lines represent the exact results for each lattice.
(b) In the case of the trellis lattice (T5),
$S_0$ is proportional to the linear size $L$ because the long-range-order
appear only in one direction. One of the ground state is shown in the inset.}
\label{entropy}
\end{figure}

To investigate more about the ground state, we investigated the residual entropy $S_0$
of the Archimedean lattices. It was calculated directly from the DOS ($\rho(E)$) obtained
by the Wang-Landau algorithm: $S_0 = \log[\rho(E_0)]$, where $E_0$ is the lowest energy.
The normalization of $\sum_{i} \rho(E_i) = 2^{N}$ should be done before the calculation.
The residual entropy was confirmed to be $S_0 = \log(2)$
for the ordered ground states (T2, T3, T10, T11, and T6), as expected.
The degeneracy 2 is from the Kramers degeneracy theorem.

As for strongly frustrated lattices (T1, T4, T7, T8, and T9),
the residual entropy is proportional to the number of lattice points $N$, 
which can be understood by applying Pauling's method to the ice model \cite{Pauling35}.
Ice can be considered as a network of corner-sharing tetrahedrons, which are composed of
an oxygen atom in the center and four hydrogen atoms at the corners. The hydrogen atom
has two possible positions (inside and outside of the tetrahedron), but should
follow the ice-rule, two-in-two-out for each tetrahedron at zero temperature.
Therefore, the residual entropy of ice can be calculated by
$S_0^{\mathrm{ice}} = \log\left[2^N (6/16)^{N/2}\right]=(N/2)\log(3/2)$,
where $2^N$ is the total number of states and $N/2$ is the number of tetrahedra. $(6/16)$ means
that only 6 states out of 16 satisfy the ice-rule for a tetrahedron.
This result is not exact, but only 1.4\% lower than recent estimates
by the Monte-Carlo method \cite{Berg07} and the series-expansion \cite{Singh12}.
If we apply the same method to the triangular and the kagom\'{e} lattices, their residual entropies become
$S_0^{\rm T1} = \log\left[2^N (6/8)^{2N}\right] \approx 0.11778 N$ and
$S_0^{\rm T8} =\log\left[2^N (6/8)^{2N/3}\right] \approx 0.50136 N$, respectively.
Since each triangle should satisfy the 2-1-rule (2 up and 1 down spins or 2 down and 1 up spins) to minimize the energy,
only 6 states out of 8 are possible in the ground state. 
This method is very close to the exact result for the kagom\'{e} lattice
($S_0^{\rm T8}/N \approx 0.50183$) \cite{Kano53},
but fails in the triangular lattice, where the exact result is
$S_0^{\rm T1}/N \approx 0.32307$ \cite{Wannier50}.
Maple leaf (T4) and bounce (T7) lattices have a unit cell with 6 points
that compose a hexagon, which is arranged in a triangular lattice. (See Fig.~\ref{Arch_fig}.)
To minimize energy, each hexagon should have one of the two configurations,
($\uparrow\downarrow\uparrow\downarrow\uparrow\downarrow$)
and ($\downarrow\uparrow\downarrow\uparrow\downarrow\uparrow$).
Since energy is lower when the nearest-neighboring hexagons
have different kind of configuration, if we define the first
kind of hexagon as spin ``up'' and the second, ``down'', the interaction
between hexagons is ``antiferromagnetic''.
Therefore, they can be regarded as a triangular Ising antiferromagnet
composed of hexagons, and have the residual entropy of $S_0^{\rm T1}$ per hexagon or
$S_0^{\rm T4,T7}/N = S_0^{\rm T1}/6N \approx 0.05384$ per lattice point.
As its nickname (expanded kagom\'{e} lattice) implies, the star lattice (T9) 
is a kagom\'{e} lattice of dimers with
spin $(\uparrow\downarrow)$ or $(\downarrow\uparrow)$.
Thus, its residual entropy can be obtained by $S_0^{\rm T9}/N = S_0^{\rm T8}/2N \approx 0.250916$.
As shown in Fig.~\ref{entropy}(a), the Wang-Landau algorithm gives consistent results
for the residual entropy of strongly frustrated lattices within 0.2\%. 

In the case of the trellis lattice, the residual entropy is proportional to the linear size $L$.
This can be understood from its ground state shown in the inset of Fig.~\ref{entropy}(b).
Spins are antiferromagnetically ordered in each row: ($\uparrow\downarrow\uparrow\downarrow \cdots$)
or ($\downarrow\uparrow\downarrow\uparrow \cdots$).
For example, when the first row has a configuration ($\uparrow\downarrow\uparrow\downarrow \cdots$),
the second row should have the configuration ($\downarrow\uparrow\downarrow\uparrow \cdots$).
Now the third row may have either configuration,
($\uparrow\downarrow\uparrow\downarrow \cdots$) or ($\downarrow\uparrow\downarrow\uparrow \cdots$).
Therefore, there exists degeneracy of 2 every other rows and the residual entropy becomes $S_0 = \log(2^L)$.
The Wang-Landau simulation supports this conclusion.
This one-dimensional long-range-ordering is possible because there exist constraints induced by
neighboring rows.


\begin{figure}[b]
\includegraphics[width=8.2cm]{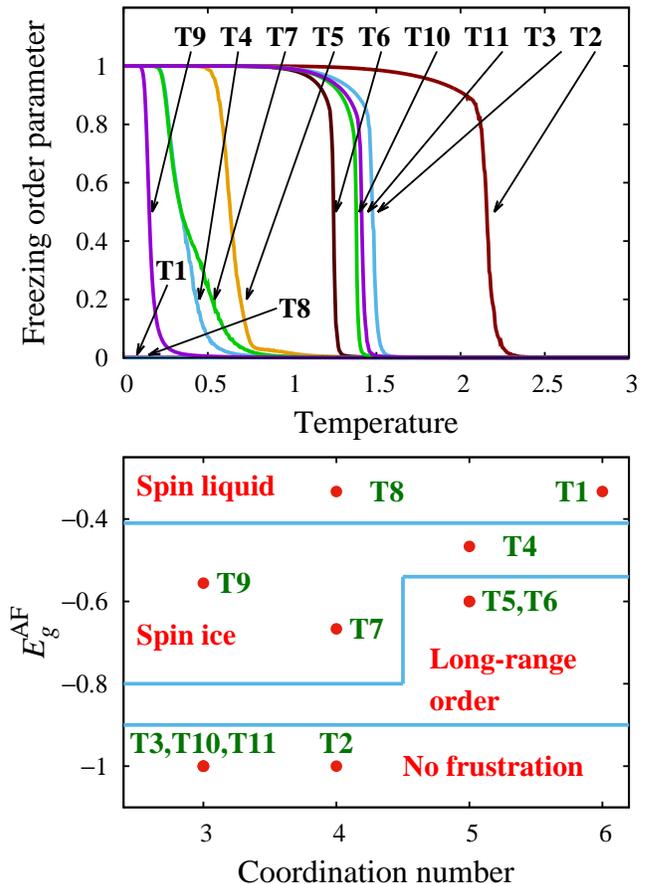}
\caption{(Color online) The freezing order parameters vs. temperature
(upper panel) and the ground state energy per bond ($E_g^{\mathrm{AF}}$) and the coordination number
of the 11 Archimedean lattices with their ground states indicated (lower panel). 
The freezing order parameters were obtained by Eq.~(\protect\ref{freezing_Eq})
with lattice size $L=40$ and number of measurements $M=2\times10^6$.
}
\label{freezing}
\end{figure}

Finally, we classified the ground state of the Ising antiferromagnet on the 11 Archimedean lattices.
Unfrustrated and weakly-frustrated lattices have a long-range-ordered ground state for
their residual entropy per site goes to zero.
The trellis lattice is a special case where the long-range-order is only in one direction.
Strongly frustrated lattices have extensive residual entropy and their ground states are disordered even at zero kelvin.
There are three possible phases: spin glass, spin ice, and spin liquid.
Since there is no disorder in lattice or magnetic interaction, which is required for the spin glass,
they are either spin ice or spin liquid.
The difference between spin ice and spin liquid comes from
their flexibility or dynamic property: The spin ice is frozen and the spin liquid fluctuates.
In order to study the fluctuation behavior, we performed a single-site-update Metropolis simulation
with $L=40$. We calculated the Edwards-Anderson order parameter $q_{\mathrm{EA}}$ \cite{EA} to measure degree of freezing.
It was proposed for the order parameter of spin glasses, but it can be used generally to study
freezing phenomena even in ferromagnets and antiferromagnets.
It can be defined in a few ways \cite{Binder86,Wengel96,Krawczyk05,Ngo14}, which are equivalent to one another.
We adopted 
\begin{eqnarray}
q_{\mathrm{EA}} = \frac{1}{N M} \sum_{i}^{N} \left| \sum_{t}^{M} S_i(t) \right| \label{freezing_Eq}
\end{eqnarray}
from Ref.~\onlinecite{Ngo14},
where $M$ is the number of measurements after equilibration, which is fixed to be $M=2\times10^6$ in this calculation.
The spins begin to freeze at the freezing temperature $T_f$, where the freezing order parameter deviates from zero.
$q_{\mathrm{EA}}$ does not depend on $M$ for $M \geq 1\times10^6$ except very close to the freezing temperature $T_f$,
where larger $M$ makes the freezing sharper.
$T_f$ are in the order of T2, T3, T11, T10, T6, T5, T7, T4, and T9.
The former 4 lattices are unfrustrated and the next two are weakly frustrated
with long-range-ordered ground states, as expected.
The last three lattices can be classified as a spin ice because the spins are frozen
without long-range-order at zero temperature.
As for T1 and T8, the freezing order parameters saturate to very small values, which decrease as $1/\sqrt{M}$.
It means spins fluctuate actively even at zero kelvin and so spin liquid phase.

These results are consistent with the theory about the degree of frustration in Ref.~\onlinecite{Richter04}.
They proposed two parameters to determine the degree of frustration: ground state energy per
bond $E_g^{\mathrm{AF}}$ and the coordination number $z$. $E_g^{\mathrm{AF}}=-1$ for unfrustrated lattices
and larger $E_g^{\mathrm{AF}}$ means more frustration. Smaller $z$ also strengthen the frustration effect.
This is also consistent with the residual entropy: Larger residual entropy implies more frustration.
Summarizing all these, the degree of frustration was determined for 7 frustrated cases
as shown in Fig.~\ref{freezing} and Table~\ref{Arch_table}.

\section{Conclusions}

In summary, we studied systematically the frustration effect of the Ising antiferromagnet
on the Archimedean lattices. From the results of specific heat, exact residual entropy, and
freezing order parameter we determined ground states of frustrated lattices.
They can be listed in the order of degree of frustration:
Shastry-Sutherland lattice and the trellis lattice (long-range-order);
bounce, maple leaf, and star lattices (spin ice phase);
triangular and kagom\'e lattices (spin liquid).

\section*{Acknowledgments}
This work was supported by the National Research Foundation Grant funded by the Korean Government (NRF-2011-0013866).

\bibliography{arch}

\begin{thebibliography}{31}
\expandafter\ifx\csname natexlab\endcsname\relax\def\natexlab#1{#1}\fi
\expandafter\ifx\csname bibnamefont\endcsname\relax
  \def\bibnamefont#1{#1}\fi
\expandafter\ifx\csname bibfnamefont\endcsname\relax
  \def\bibfnamefont#1{#1}\fi
\expandafter\ifx\csname citenamefont\endcsname\relax
  \def\citenamefont#1{#1}\fi
\expandafter\ifx\csname url\endcsname\relax
  \def\url#1{\texttt{#1}}\fi
\expandafter\ifx\csname urlprefix\endcsname\relax\def\urlprefix{URL }\fi
\providecommand{\bibinfo}[2]{#2}
\providecommand{\eprint}[2][]{\url{#2}}

\bibitem[{\citenamefont{Ising}(1925)}]{Ising25}
\bibinfo{author}{\bibfnamefont{E.}~\bibnamefont{Ising}}, \bibinfo{journal}{Z.
  Phys.} \textbf{\bibinfo{volume}{31}}, \bibinfo{pages}{253}
  (\bibinfo{year}{1925}).

\bibitem[{\citenamefont{Peierls}(1936)}]{Peierls36}
\bibinfo{author}{\bibfnamefont{R.~E.} \bibnamefont{Peierls}},
  \bibinfo{journal}{Proc. Cambridge Philos. Soc.}
  \textbf{\bibinfo{volume}{32}}, \bibinfo{pages}{477} (\bibinfo{year}{1936}).

\bibitem[{\citenamefont{Onsager}(1944)}]{Onsager44}
\bibinfo{author}{\bibfnamefont{L.}~\bibnamefont{Onsager}},
  \bibinfo{journal}{Phys. Rev.} \textbf{\bibinfo{volume}{65}},
  \bibinfo{pages}{117} (\bibinfo{year}{1944}).

\bibitem[{\citenamefont{Wannier}(1950)}]{Wannier50}
\bibinfo{author}{\bibfnamefont{G.~H.} \bibnamefont{Wannier}},
  \bibinfo{journal}{Phys. Rev.} \textbf{\bibinfo{volume}{79}},
  \bibinfo{pages}{357} (\bibinfo{year}{1950}).

\bibitem[{\citenamefont{Kan\^{o} and Naya}(1953)}]{Kano53}
\bibinfo{author}{\bibfnamefont{K.}~\bibnamefont{Kan\^{o}}} \bibnamefont{and}
  \bibinfo{author}{\bibfnamefont{S.}~\bibnamefont{Naya}},
  \bibinfo{journal}{Prog. Theo. Phys.} \textbf{\bibinfo{volume}{10}},
  \bibinfo{pages}{158} (\bibinfo{year}{1953}).

\bibitem[{\citenamefont{Balents}(2010)}]{Balents10}
\bibinfo{author}{\bibfnamefont{L.}~\bibnamefont{Balents}},
  \bibinfo{journal}{Nature} \textbf{\bibinfo{volume}{464}},
  \bibinfo{pages}{199} (\bibinfo{year}{2010}).

\bibitem[{\citenamefont{Diep}(2013)}]{Diep}
\bibinfo{author}{\bibfnamefont{H.~T.} \bibnamefont{Diep}},
  \emph{\bibinfo{title}{Frustrated Spin Systems}} (\bibinfo{publisher}{World
  Scientific}, \bibinfo{address}{Singapore}, \bibinfo{year}{2013}),
  \bibinfo{edition}{2nd} ed.

\bibitem[{\citenamefont{Ramirez}(1994)}]{Ramirez94}
\bibinfo{author}{\bibfnamefont{A.~P.} \bibnamefont{Ramirez}},
  \bibinfo{journal}{Annu. Rev. Mater. Sci.} \textbf{\bibinfo{volume}{24}},
  \bibinfo{pages}{453} (\bibinfo{year}{1994}).

\bibitem[{\citenamefont{Gr\"{u}nbaum and Shephard}(1987)}]{Grunbaum87}
\bibinfo{author}{\bibfnamefont{B.}~\bibnamefont{Gr\"{u}nbaum}}
  \bibnamefont{and} \bibinfo{author}{\bibfnamefont{G.~C.}
  \bibnamefont{Shephard}}, \emph{\bibinfo{title}{Tilings and Patterns}}
  (\bibinfo{publisher}{W. H. Freeman and Company}, \bibinfo{address}{New York},
  \bibinfo{year}{1987}).

\bibitem[{\citenamefont{Richter et~al.}(2004)\citenamefont{Richter,
  Schulenburg, and Honecker}}]{Richter04}
\bibinfo{author}{\bibfnamefont{J.}~\bibnamefont{Richter}},
  \bibinfo{author}{\bibfnamefont{J.}~\bibnamefont{Schulenburg}},
  \bibnamefont{and} \bibinfo{author}{\bibfnamefont{A.}~\bibnamefont{Honecker}},
  in \emph{\bibinfo{booktitle}{Lecture Notes in Physics}}, edited by
  \bibinfo{editor}{\bibfnamefont{U.}~\bibnamefont{Schollw\"{o}ck}},
  \bibinfo{editor}{\bibfnamefont{J.}~\bibnamefont{Richter}},
  \bibinfo{editor}{\bibfnamefont{D.~J.~J.} \bibnamefont{Farnell}},
  \bibnamefont{and} \bibinfo{editor}{\bibfnamefont{R.~F.} \bibnamefont{Bishop}}
  (\bibinfo{publisher}{Springer}, \bibinfo{address}{Berlin},
  \bibinfo{year}{2004}), vol. \bibinfo{volume}{645}.

\bibitem[{\citenamefont{Zheng et~al.}(2014)\citenamefont{Zheng, Zheng, and
  Chen}}]{Zheng14}
\bibinfo{author}{\bibfnamefont{Y.~Z.} \bibnamefont{Zheng}},
  \bibinfo{author}{\bibfnamefont{Z.}~\bibnamefont{Zheng}}, \bibnamefont{and}
  \bibinfo{author}{\bibfnamefont{X.~M.} \bibnamefont{Chen}},
  \bibinfo{journal}{Coord. Chem. Rev.} \textbf{\bibinfo{volume}{258-259}},
  \bibinfo{pages}{1} (\bibinfo{year}{2014}).

\bibitem[{\citenamefont{Suding and Ziff}(1999)}]{Suding99}
\bibinfo{author}{\bibfnamefont{P.~N.} \bibnamefont{Suding}} \bibnamefont{and}
  \bibinfo{author}{\bibfnamefont{R.~M.} \bibnamefont{Ziff}},
  \bibinfo{journal}{Phys. Rev. E} \textbf{\bibinfo{volume}{60}},
  \bibinfo{pages}{275} (\bibinfo{year}{1999}).

\bibitem[{\citenamefont{Neher et~al.}(2008)\citenamefont{Neher, Mecke, and
  Wagner}}]{Neher08}
\bibinfo{author}{\bibfnamefont{R.~A.} \bibnamefont{Neher}},
  \bibinfo{author}{\bibfnamefont{K.}~\bibnamefont{Mecke}}, \bibnamefont{and}
  \bibinfo{author}{\bibfnamefont{H.}~\bibnamefont{Wagner}},
  \bibinfo{journal}{J. Stat. Mech.} \textbf{\bibinfo{volume}{2008}},
  \bibinfo{pages}{P01011} (\bibinfo{year}{2008}).

\bibitem[{\citenamefont{Codello}(2010)}]{Codello10}
\bibinfo{author}{\bibfnamefont{A.}~\bibnamefont{Codello}}, \bibinfo{journal}{J.
  Phys. A} \textbf{\bibinfo{volume}{43}}, \bibinfo{pages}{385002}
  (\bibinfo{year}{2010}).

\bibitem[{\citenamefont{Farnell et~al.}(2014)\citenamefont{Farnell, G\"{o}tze,
  Richter, Bishop, and Li}}]{Farnell14}
\bibinfo{author}{\bibfnamefont{D.~J.~J.} \bibnamefont{Farnell}},
  \bibinfo{author}{\bibfnamefont{O.}~\bibnamefont{G\"{o}tze}},
  \bibinfo{author}{\bibfnamefont{J.}~\bibnamefont{Richter}},
  \bibinfo{author}{\bibfnamefont{R.~F.} \bibnamefont{Bishop}},
  \bibnamefont{and} \bibinfo{author}{\bibfnamefont{P.~H.~Y.} \bibnamefont{Li}},
  \bibinfo{journal}{Phys. Rev. B} \textbf{\bibinfo{volume}{89}},
  \bibinfo{pages}{184407} (\bibinfo{year}{2014}).

\bibitem[{\citenamefont{Shastry and Sutherland}(1981)}]{SSL}
\bibinfo{author}{\bibfnamefont{B.}~\bibnamefont{Shastry}} \bibnamefont{and}
  \bibinfo{author}{\bibfnamefont{B.}~\bibnamefont{Sutherland}},
  \bibinfo{journal}{Physica B} \textbf{\bibinfo{volume}{108}},
  \bibinfo{pages}{1069} (\bibinfo{year}{1981}).

\bibitem[{\citenamefont{Wolff}(1989)}]{Wolff89}
\bibinfo{author}{\bibfnamefont{U.}~\bibnamefont{Wolff}},
  \bibinfo{journal}{Phys. Rev. Lett.} \textbf{\bibinfo{volume}{62}},
  \bibinfo{pages}{361} (\bibinfo{year}{1989}).

\bibitem[{\citenamefont{Ferrenberg and Swendsen}(1988)}]{Ferrenberg88}
\bibinfo{author}{\bibfnamefont{A.~M.} \bibnamefont{Ferrenberg}}
  \bibnamefont{and} \bibinfo{author}{\bibfnamefont{R.~H.}
  \bibnamefont{Swendsen}}, \bibinfo{journal}{Phys. Rev. Lett.}
  \textbf{\bibinfo{volume}{61}}, \bibinfo{pages}{2635} (\bibinfo{year}{1988}).

\bibitem[{\citenamefont{Ferrenberg and Swendsen}(1989)}]{Ferrenberg89}
\bibinfo{author}{\bibfnamefont{A.~M.} \bibnamefont{Ferrenberg}}
  \bibnamefont{and} \bibinfo{author}{\bibfnamefont{R.~H.}
  \bibnamefont{Swendsen}}, \bibinfo{journal}{Phys. Rev. Lett.}
  \textbf{\bibinfo{volume}{63}}, \bibinfo{pages}{1195} (\bibinfo{year}{1989}).

\bibitem[{\citenamefont{Yu}(2015)}]{Yu15}
\bibinfo{author}{\bibfnamefont{U.}~\bibnamefont{Yu}}, \bibinfo{journal}{Physica
  A} \textbf{\bibinfo{volume}{419}}, \bibinfo{pages}{75}
  (\bibinfo{year}{2015}).

\bibitem[{\citenamefont{Wang and Landau}(2001)}]{Wang01}
\bibinfo{author}{\bibfnamefont{F.}~\bibnamefont{Wang}} \bibnamefont{and}
  \bibinfo{author}{\bibfnamefont{D.~P.} \bibnamefont{Landau}},
  \bibinfo{journal}{Phys. Rev. Lett.} \textbf{\bibinfo{volume}{86}},
  \bibinfo{pages}{2050} (\bibinfo{year}{2001}).

\bibitem[{\citenamefont{Metropolis et~al.}(1953)\citenamefont{Metropolis,
  Rosenbluth, Rosenbluth, Teller, and Teller}}]{Metropolis53}
\bibinfo{author}{\bibfnamefont{N.}~\bibnamefont{Metropolis}},
  \bibinfo{author}{\bibfnamefont{A.~W.} \bibnamefont{Rosenbluth}},
  \bibinfo{author}{\bibfnamefont{M.~N.} \bibnamefont{Rosenbluth}},
  \bibinfo{author}{\bibfnamefont{A.~M.} \bibnamefont{Teller}},
  \bibnamefont{and} \bibinfo{author}{\bibfnamefont{E.}~\bibnamefont{Teller}},
  \bibinfo{journal}{J. Chem. Phys.} \textbf{\bibinfo{volume}{21}},
  \bibinfo{pages}{1087} (\bibinfo{year}{1953}).

\bibitem[{\citenamefont{Barber}(1983)}]{FSS}
\bibinfo{author}{\bibfnamefont{M.~N.} \bibnamefont{Barber}}, in
  \emph{\bibinfo{booktitle}{Phase Transitions and Critical Phenomena}}, edited
  by \bibinfo{editor}{\bibfnamefont{C.}~\bibnamefont{Domb}} \bibnamefont{and}
  \bibinfo{editor}{\bibfnamefont{J.}~\bibnamefont{Lebowitz}}
  (\bibinfo{publisher}{Academic}, \bibinfo{address}{New York},
  \bibinfo{year}{1983}), vol.~\bibinfo{volume}{8}.

\bibitem[{\citenamefont{Pauling}(1935)}]{Pauling35}
\bibinfo{author}{\bibfnamefont{L.}~\bibnamefont{Pauling}}, \bibinfo{journal}{J.
  Am. Chem. Soc.} \textbf{\bibinfo{volume}{57}}, \bibinfo{pages}{2680}
  (\bibinfo{year}{1935}).

\bibitem[{\citenamefont{Berg et~al.}(2007)\citenamefont{Berg, Muguruma, and
  Okamoto}}]{Berg07}
\bibinfo{author}{\bibfnamefont{B.~A.} \bibnamefont{Berg}},
  \bibinfo{author}{\bibfnamefont{C.}~\bibnamefont{Muguruma}}, \bibnamefont{and}
  \bibinfo{author}{\bibfnamefont{Y.}~\bibnamefont{Okamoto}},
  \bibinfo{journal}{Phys. Rev. B} \textbf{\bibinfo{volume}{75}},
  \bibinfo{pages}{092202} (\bibinfo{year}{2007}).

\bibitem[{\citenamefont{Singh and Oitmaa}(2012)}]{Singh12}
\bibinfo{author}{\bibfnamefont{R.~R.~P.} \bibnamefont{Singh}} \bibnamefont{and}
  \bibinfo{author}{\bibfnamefont{J.}~\bibnamefont{Oitmaa}},
  \bibinfo{journal}{Phys. Rev. B} \textbf{\bibinfo{volume}{85}},
  \bibinfo{pages}{144414} (\bibinfo{year}{2012}).

\bibitem[{\citenamefont{Edwards and Anderson}(1975)}]{EA}
\bibinfo{author}{\bibfnamefont{S.~F.} \bibnamefont{Edwards}} \bibnamefont{and}
  \bibinfo{author}{\bibfnamefont{P.}~\bibnamefont{Anderson}},
  \bibinfo{journal}{J. Phys. F: Metal Phys.} \textbf{\bibinfo{volume}{5}},
  \bibinfo{pages}{965} (\bibinfo{year}{1975}).

\bibitem[{\citenamefont{Wengel et~al.}(1996)\citenamefont{Wengel, Henley, and
  Zippelius}}]{Wengel96}
\bibinfo{author}{\bibfnamefont{C.}~\bibnamefont{Wengel}},
  \bibinfo{author}{\bibfnamefont{C.~L.} \bibnamefont{Henley}},
  \bibnamefont{and}
  \bibinfo{author}{\bibfnamefont{A.}~\bibnamefont{Zippelius}},
  \bibinfo{journal}{Phys. Rev. B} \textbf{\bibinfo{volume}{53}},
  \bibinfo{pages}{6543} (\bibinfo{year}{1996}).

\bibitem[{\citenamefont{Krawczyk et~al.}(2005)\citenamefont{Krawczyk, Malarz,
  Kawecka-Magiera, Maksymowicz, and Ku\l{}akowski}}]{Krawczyk05}
\bibinfo{author}{\bibfnamefont{M.~J.} \bibnamefont{Krawczyk}},
  \bibinfo{author}{\bibfnamefont{K.}~\bibnamefont{Malarz}},
  \bibinfo{author}{\bibfnamefont{B.}~\bibnamefont{Kawecka-Magiera}},
  \bibinfo{author}{\bibfnamefont{A.~Z.} \bibnamefont{Maksymowicz}},
  \bibnamefont{and}
  \bibinfo{author}{\bibfnamefont{K.}~\bibnamefont{Ku\l{}akowski}},
  \bibinfo{journal}{Phys. Rev. B} \textbf{\bibinfo{volume}{72}},
  \bibinfo{pages}{024445} (\bibinfo{year}{2005}).

\bibitem[{\citenamefont{Ngo et~al.}(2014)\citenamefont{Ngo, Hoang, Diep, and
  Campbell}}]{Ngo14}
\bibinfo{author}{\bibfnamefont{V.~T.} \bibnamefont{Ngo}},
  \bibinfo{author}{\bibfnamefont{D.~T.} \bibnamefont{Hoang}},
  \bibinfo{author}{\bibfnamefont{H.~T.} \bibnamefont{Diep}}, \bibnamefont{and}
  \bibinfo{author}{\bibfnamefont{I.~A.} \bibnamefont{Campbell}},
  \bibinfo{journal}{Mod. Phys. Lett. B} \textbf{\bibinfo{volume}{28}},
  \bibinfo{pages}{1450067} (\bibinfo{year}{2014}).

\bibitem[{\citenamefont{Binder and Young}(1986)}]{Binder86}
\bibinfo{author}{\bibfnamefont{K.}~\bibnamefont{Binder}} \bibnamefont{and}
  \bibinfo{author}{\bibfnamefont{A.~P.} \bibnamefont{Young}},
  \bibinfo{journal}{Rev. Mod. Phys.} \textbf{\bibinfo{volume}{58}},
  \bibinfo{pages}{801} (\bibinfo{year}{1986}).

\end{thebibliography}

\end{document}